\documentclass[12pt,twoside]{iopart}

\usepackage{iopams} 
\usepackage{graphicx, subfigure}
\usepackage{balance}
\usepackage{bm}
\usepackage{float}
\usepackage{color}
\newcommand{\ex}[1]{\times10^{#1}}

\begin{document}

\title[]{Confined Cubic Blue Phases under Shear}
\author{O. Henrich$^{1}$, K. Stratford$^2$, D. Marenduzzo$^3$, P.V. Coveney$^1$ and M.E. Cates $^3$}

\address{$^1$ Centre for Computational Science, Department of Chemistry,
University College London, WC1H 0AJ, United Kingdom\\
$^2$ Edinburgh Parallel Computing Centre and
$^3$ School of Physics and Astronomy, The King's Buildings,
The University of Edinburgh, EH9 3JZ, United Kingdom}
\ead{o.henrich@ucl.ac.uk}

\begin{abstract}
We study the behaviour of confined cubic blue phases under shear flow
via lattice Boltzmann simulations. We focus on the two experimentally observed
phases, blue phase I and blue phase II. The disclination network of blue phase II continuously breaks and reforms under shear, 
leading to an oscillatory stress response in time. The oscillations are only regular for very thin samples. For thicker samples, the shear leads to a ``stick-slip'' motion
of part of the network along the vorticity direction.
Blue phase I responds very differently: its defect network undergoes seemingly
chaotic rearrangements under shear, irrespective of system size. 
\end{abstract}


\section{Introduction}
Cholesterics are liquid crystals in which the local nematic director field spontaneously twists in thermodynamic equilibrium~\cite{deGennes}.
The preferred configuration close to the isotropic boundary features twist around two perpendicular axes, as opposed to just one axis in the regular cholesteric state, and the corresponding deformation is denoted a ``double-twist cylinder''.
As it is topologically impossible to cover continuously 3D space with double-twist cylinders, defects arise. These organise into a variety of regular periodic lattices, giving rise to the so-called cubic blue phases (BPs) \cite{Grebel:1984,Wright:1989}. There are two experimentally observed cubic blue phases, BPI and BPII (a third, BPIII, is thought to be amorphous~\cite{Henrich:2011a}).

BPs have been long considered as purely of academic interest due to their very narrow range of stability. This view has changed since the creation of
polymer-stabilised and temperature-stabilised BPs~\cite{Kikuchi:2002,Coles:2005}, which has opened up the possibility of novel applications.
During the last few years considerable progress has been achieved regarding the behaviour of BPs in confined geometries~\cite{Fukuda:2010a, Fukuda:2010b, Ravnik:2011b}, under external fields \cite{Alexander:2008,Fukuda:2009,Henrich:2010a,Castles:2010,Tiribocchi:2011}, and in the presence of colloidal particles \cite{Ravnik:2011a}. 
The kinetics of BP domain growth have been recently addressed in~\cite{Henrich:2010b}. However, our understanding of their dynamical behaviour under flow remains
very limited. The aim of this work is to address this issue by studying,
for the first time, the response of confined BP samples to a shear flow.

Flow response in cholesterics is both strongly non-Newtonian and highly anisotropic.
For example, if a cholesteric helix is subjected to a Poiseuille flow along
the helical axis, small pressure differences drive flow mainly through
``permeation'', first investigated by Helfrich \cite{Helfrich:1969}.
In the permeation mode the liquid crystal flows while leaving the director
field virtually unchanged, which leads to high dissipation and large
viscosities. Marenduzzo et al. \cite{Marenduzzo:2006a,Marenduzzo:2006b} simulated shear and Poiseuille flow in cholesteric liquid crystals in the permeation mode, and showed the importance of the boundary conditions in determining the apparent viscosity of the fluid. They also found that a strong secondary flow appears.
Rey \cite{Rey:1996a, Rey:1996b} studied shear in cholesterics oriented with the helix along the vorticity axis and found that, at low Ericksen number, travelling twist waves appear which lead to the rotation of the cholesteric helix. At higher forcing, the helix uncoils and leaves a flow-induced nematic phase. 
Rey also studied cholesterics subjected to both steady flow and low frequency
small amplitude oscillatory shear for different helix orientations
\cite{Rey:2000, Rey:2002}. He found that splay/bend/twist deformations were
excited when the helix was aligned along the flow direction; splay/bend
deformation occurred when the helix was aligned along the velocity gradient;
but only twist deformations
appeared when the helix was aligned along the vorticity axis.

Dupuis et al. \cite{Dupuis:2005} performed the first numerical investigation of BP rheology in Poiseuille flow, starting from equilibrium structures of BPI and BPII and a periodic array of doubly twisted cylinders.
Under small forcing, the network opposed the flow giving rise to a significant increase in apparent viscosity.
Upon increasing the forcing they found clear evidence of shear thinning.
In the crossover region they predicted a novel oscillatory regime where the network continuously breaks and reforms as portions of the disclinations in the center of the channel move to neighboring cells and relink with the parts of the network left behind by the flow. Compared with the cholesteric case, the viscosity still decreases with forcing (the system shear thins) but much less than
for cholesterics in the permeation mode, which is in agreement with experiments
\cite{Zapotocky:1999, Ramos:2002}.

The purpose of the present paper is to contribute to the understanding of the 
flow behaviour of cholesteric blue phases by reporting results of the first
supra-unit cell simulations. This paper is organised as follows.
In Sec.~\ref{sec_model} we introduce the Beris-Edwards model for hydrodynamics
of liquid crystals, which is a generalisation of Ericksen's and Leslie's
theory of nematodynamics~\cite{deGennes}.
In Sec.~\ref{sec_results} we describe the simulation methodology and the
boundary conditions we applied.
Results for a confined BPII in rectilinear shear flow are presented in
Sec.~\ref{sec_bpii} with particular focus on the dependence of the flow
on the thickness of the sample and thermodynamic state.
In Sec.~\ref{sec_bpi} we show corresponding results for BPI, which indicate 
fundamental differences in the flow behaviour of BPI and BPII. 

\section{Model}\label{sec_model}

Our approach is based on the well-established Beris-Edwards model for hydrodynamics of
cholesteric liquid crystals \cite{Beris:1994}, which describes the ordered state 
in terms of a traceless, symmetric tensor order parameter ${\mathbf Q}({\mathbf r})$. 
In the uniaxial approximation, the order parameter is given by
$Q_{\alpha \beta}= q_s ( n_\alpha n_\beta - \frac{1}{3}\; \delta_{\alpha\beta})$
with ${\mathbf n}$ the director field and $q_s$ the amplitude of nematic
order. More generally,
the largest eigenvalue of ${\mathbf Q}$, $0\le q_s\le\frac{2}{3}$
characterizes the local degree of orientational order.
The thermodynamic properties of the liquid crystal are determined by a free energy
${\cal F}$, whose density $f$ consists of a bulk contribution $f_b$ and a gradient part $f_g$, as follows,
\begin{eqnarray}
f_b&=&\frac{A_0}{2}\left(1-\frac{\gamma}{3}\right) Q_{\alpha \beta}^2-\frac{A_0 \gamma}{3}Q_{\alpha \beta} Q_{\beta \gamma} Q_{\gamma \alpha}+\frac{A_0 \gamma}{4}(Q_{\alpha \beta}^2)^2,\nonumber\\
f_g&=&\frac{K}{2}(\varepsilon_{\alpha\gamma\delta} \partial_\gamma Q_{\delta\beta}+2 q_0 Q_{\alpha \beta})^2+\frac{K}{2}(\partial_\beta Q_{\alpha \beta})^2.\label{FE}
\end{eqnarray}
The first term contains the bulk-free energy constant $A_0$ and the inverse temperature $\gamma$ which controls the magnitude of order.
The second part quantifies the cost of elastic distortions, which are proportional to the elastic constant $K$;
we work for simplicity in the one-elastic constant approximation. The wavevector $q_0=2\pi/p_0$, where $p_0$ is the cholesteric pitch.
The actual periodicity of the BP structure, $p$, does not need to be equal to $p_0$.
The ``redshift'' $r=p/p_0$ is adjusted during the simulation by following a simple procedure similar to \cite{Alexander:2006}.

A particular thermodynamic state is specified by two dimensionless quantities: the effective temperature $\tau$ and chirality $\kappa$,
which are given by
\begin{eqnarray}
\tau&=&\frac{27(1-\gamma/3)}{\gamma}\nonumber\\
\kappa&=&\sqrt{\frac{108 K q_0^2}{A_0 \gamma}}\nonumber.
\end{eqnarray}
The dynamical evolution of the order parameter is given by the equation 
\begin{equation}
\left(\partial_t+ u_\alpha \partial_\alpha \right){\mathbf Q} - {\mathbf S}({\mathbf W},{\mathbf Q}) = \Gamma {\mathbf H}.
\label{eqn2}
\end{equation}
The first term on the left hand side of Eq.\ref{eqn2} is a material derivative, which describes the rate of change of a quantity moving along with the flow.
The second term accounts for the rate of change due to local velocity gradients $W_{\alpha \beta}=\partial_\beta u_\alpha$,
\begin{eqnarray}
{\mathbf S}({\mathbf W}, {\mathbf Q}) &=& (\xi {\mathbf A} + {\boldsymbol \Omega})({\mathbf Q}+\frac{\mathbf I}{3})\nonumber\\
& &\hspace*{-1.5cm}+ ({\mathbf Q}+\frac{\mathbf I}{3})(\xi {\mathbf A}  - {\boldsymbol \Omega})-2 \xi ({\mathbf Q}+\frac{\mathbf I}{3})
\mathrm{Tr}({\mathbf Q W}),
\label{eqn3}
\end{eqnarray}
where $\mathrm{Tr}$ denotes the tensorial trace, while 
${\mathbf A}=({\mathbf W}+{\mathbf W}^T)/2$ and
${\boldsymbol \Omega}=({\mathbf W}-{\mathbf W}^T)/2$ are the symmetric and antisymmetric part of the velocity gradient, respectively. $\xi$ is a constant depending on the molecular details of the liquid crystal.
Flow alignment occurs if $\xi \cos{2\theta}=(3q_s)/(2+q_s)$ has a real solution, where $\theta$ is the Leslie-angle: we select this case by setting $\xi=0.7$ in our simulations.
${\mathbf H}$ is the molecular field, which is a functional derivative of $\cal F$ that preserves the tracelessness of $\mathbf Q$:
\begin{equation}
{\bf H}=-\frac{\delta {\cal F}}{\delta {\bf Q}}+\frac{\bf I}{3}\,
\mathrm{Tr} \left(\frac{\delta {\cal F}}{\delta {\bf Q}}\right).
\label{eqn4}
\end{equation}
The rotational diffusion constant $\Gamma$ in Eq.~\ref{eqn2} is proportional
to the inverse of the rotational viscosity $\gamma_1=2 q_s^2/\Gamma$
\cite{deGennes}.

The time evolution of the fluid density and velocity are respectively governed
by the continuity equation
$\partial_t \rho = -\partial_\alpha(\rho u_\alpha)$, and
the following Navier-Stokes equation:
\begin{eqnarray}
\partial_t u_\alpha +\rho \,u_\beta \partial_\beta u_\alpha
= \partial_\beta \Pi_{\alpha \beta}
+\eta\, \partial_\beta \{ \partial_\alpha u_\beta + \partial_\beta u_\alpha
+(1+3\frac{\partial P_0}{\partial\rho} )\partial_\mu u_\mu \delta_{\alpha \beta}\}. 
\label{eqn6}
\end{eqnarray}
The final term in Eq.~\ref{eqn6} arises from the Chapman-Enskog expansion
of the LB equations \cite{Denniston:2001}.
At low flow rates the fluid can be considered as incompressible, so that the
last term on the right hand side of Eq.~\ref{eqn6} remains small.
$\eta$ is an isotropic background viscosity which is set to $\eta=0.8333$ in LBU.
The pressure tensor reads explicitly
\begin{eqnarray}
\Pi_{\alpha \beta}&=&P_0 \delta_{\alpha\beta}
-\xi H_{\alpha \gamma}\left(Q_{\gamma \beta} +\frac{1}{3} \delta_{\gamma \beta}\right)-\xi \left(Q_{\alpha \gamma} +\frac{1}{3} \delta_{\alpha \gamma}\right) H_{\gamma \beta}\nonumber\\
&+& 2 \xi  \left(Q_{\alpha \beta} +\frac{1}{3} \delta_{\alpha \beta}\right) Q_{\gamma \nu} H_{\gamma \nu}-\partial_\alpha Q_{\gamma \nu} \frac{\delta{\cal F}}{\delta \partial_{\beta} Q_{\gamma \nu}}\nonumber\\
&+&Q_{\alpha \gamma}H_{\gamma \beta}-H_{\alpha \gamma} Q_{\gamma \beta}.
\label{eqn7}
\end{eqnarray}
In the isotropic state ${\bf Q}\equiv 0$ and Eq.\ref{eqn7} is reduced to the
scalar pressure which, in a system at rest, is constant to a very good
approximation.
 
The system of coupled partial differential equations Eqs.~\ref{eqn2}
and~\ref{eqn6} is solved by means of a
hybrid scheme \cite{Marenduzzo:2007}. This uses a lattice Boltzmann algorithm
with predictor-corrector scheme for the continuity equation and
Eq.~\ref{eqn6}, and a finite difference scheme for the equation of motion of
the tensor order parameter Eq.\ref{eqn2}. More details on the algorithm can
be found in \cite{Denniston:2001, Denniston:2004}.

\section{Methodology and Results}\label{sec_results}

We present results from simulations of confined cubic blue
phases in steady shear flow.
Our system consists of $n_x\times n_z = 4\times4$ unit cells in $x$-
and $z$-direction (flow and vorticity direction respectively) and a varying
number $n_y$ of unit cells in $y$-direction (which is the flow gradient
direction in our geometry).
As is common in BP simulation studies ~\cite{Henrich:2011a,Henrich:2010b}, we initialised our simulations with analytical solutions that minimize the free energy functional Eq.\ref{FE} in the high-chirality limit, and equilibrated these (for 3000 LB timesteps) with periodic boundary conditions prior to starting the shear. During the equilibration run the optimal redshift $r$ was recalculated at every timestep.

Shear flow was then started by applying shear boundary conditions to the LB distribution functions as in~\cite{Denniston:2004}. For ${\mathbf Q}$, the equilibrated order parameter was advected on the top and bottom boundary, according to the imposed wall velocities. Periodic boundary conditions were retained along the flow and vorticity direction.
This set-up is similar to the one Dupuis et al. used to simulate Poiseuille flow \cite{Dupuis:2005}. The pinning of the defect network at the boundaries may be realised in practice by impurities or surface defects.
During shear flow the unit cell size was kept fixed and a constant redshift equal to the value at the end of the equilibration protocol was assumed.

The total number of unit cells in our simulations of both BPI and BPII
was $n_x\times n_y \times n_z= 4 \times \{1,2,3,4\} \times 4$.
A resolution of 32 lattice Boltzmann units (LBU) per pitch length
(16 per unit cell) was used for BPII, and 64 LBU per pitch length
(32 per unit cell) used for BPI.  
This provided enough spatial resolution to track even complex rearrangements
of the disclination networks in each case.
The timestep and lattice spacing in LBU corresponds roughly to
$\sim 1 {\rm ns}$ and $\sim 10{\rm nm}$ in SI units. The LBU of stress
is equal to about $10^8$~Pa. 
More details about the conversion from LBU to SI units can be found in
\cite{Henrich:2011a,Henrich:2010b}.

\subsection{Blue Phase II}\label{sec_bpii}

\begin{figure}[t]
\includegraphics[width=0.24\textwidth]{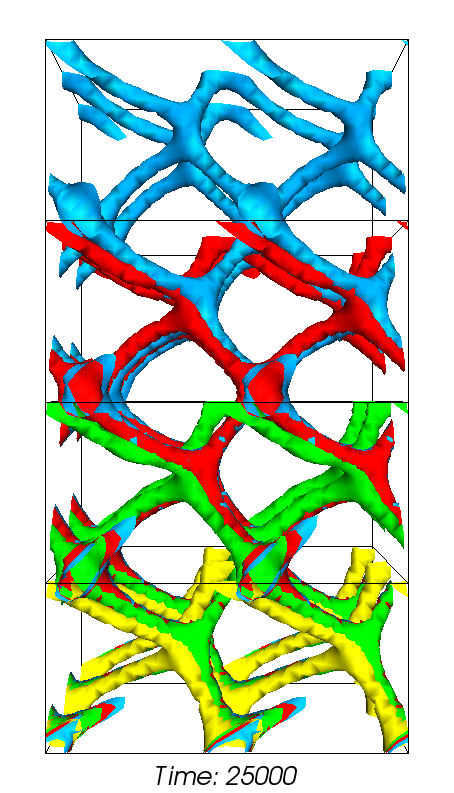}
\includegraphics[width=0.24\textwidth]{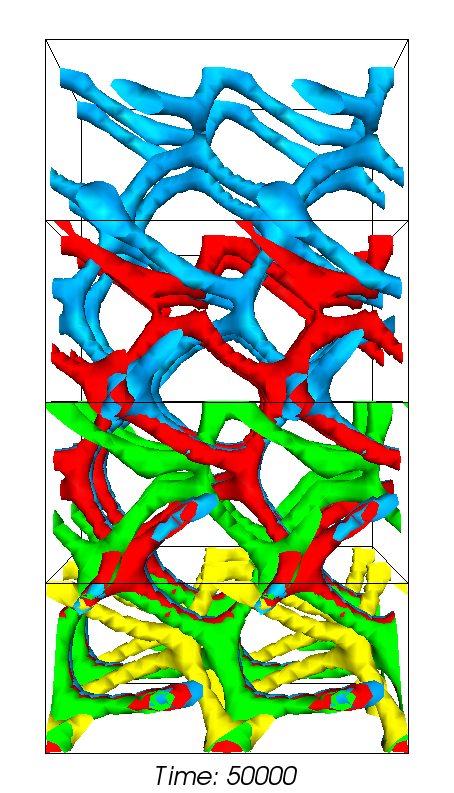}
\includegraphics[width=0.24\textwidth]{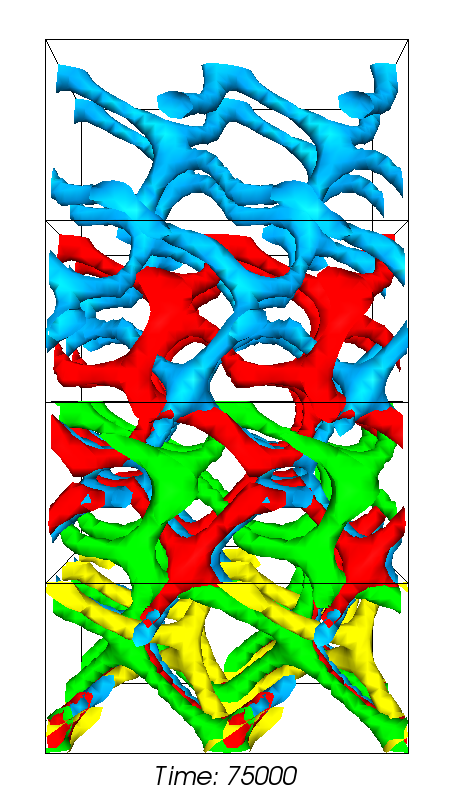}
\includegraphics[width=0.24\textwidth]{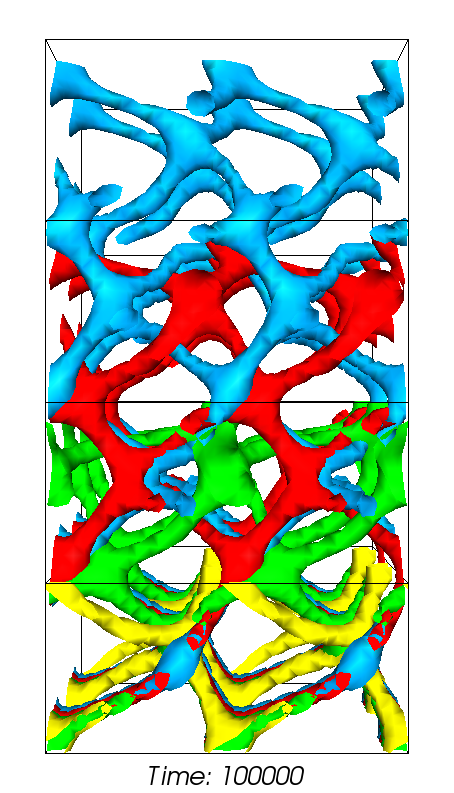}
\caption{Shear flow of BPII at different gap sizes. The images show
snapshots of a subset of the entire disclination network in a system
with walls at top and bottom, and the flow direction into/out of the
page. The different number of unit cells in the velocity gradient
direction have been superposed and are shown in different colours:
$n_y=1 \mbox{ (yellow)}, n_y=2 \mbox{ (green)}, n_y=3 \mbox{ (red) and }
n_y=4 \mbox{ (blue)}$. All cases show $n_x \times n_z =2 \times 2$ unit
cells in flow and vorticity directions. The temperature and chirality are
$\tau=-0.5$ and $\kappa=2$, respectively, where BPII is the equlibrium phase.
The strain rate is $\dot{\gamma}=7.8125\ex{-5}$ in reciprocal LB time units in all cases.}
\label{fig1}
\end{figure}

We first report the results for BPII.
Fig.~\ref{fig1} shows the disclination network in steady shear flow for different numbers of unit cells between the walls.
(We prefer the perspective along the flow direction because the network topology is clearer.)
Temperature and chirality are $\tau=-0.5$ and $\kappa=2$ respectively, and for these parameters BPII is the equilibrium phase.
The velocity of the walls has been adapted so that the same shear rate $\dot{\gamma}=7.8125\ex{-5}$ is achieved for all gap sizes.
Shear rates four times larger and smaller were also applied and produced very similar results, apart from the flow velocities which scale up accordingly.
For all gap sizes we observe an affine transformation in the gradient-flow plane that leads to a break-up of the network at strains of about half a unit cell size.
Shortly after the break-up the network reforms further downstream, resulting in a periodic, regular structure that is very close to the equilibrium
BPII disclination network. Similar oscillations have been previously reported for
BPs in Poiseuille flow, with a gap of one unit cell size \cite{Dupuis:2005}.

It is interesting to compare the results for different gap sizes.
For $n_y=1$ we observe that the BPII-network is slightly deformed with
respect to the quiescent state, but always reforms at the same position.
For $n_y > 1$, the BPII network distorts more under the same shear rate, 
suggesting that the structure within the sample is strongly affected by
the anchoring at the boundaries.
Interestingly, when the sample is thicker than one unit cell, a movement
of the network in the positive vorticity direction arises
which is not observed for $n_y=1$.
This network motion does not proceed at a constant speed, but rather
proceeds in a ``stick-slip'' fashion: elastic stresses first slowly build up,
and are then quickly released as the network ``slips'' along the vorticity
direction. Changing from a left-handed to a right-handed helix inverts the
direction of motion of the network.
Hence the ``stick-slip'' motion is related to the chiral nature of the helix.
This phenomenon may bear some similarity to the travelling wave motion in
cholesterics sheared along the vorticity direction 
below a critical Ericksen number \cite{Rey:1996a, Rey:1996b}.

\begin{figure*}[t]
\centering
\includegraphics[width=0.3\textwidth]{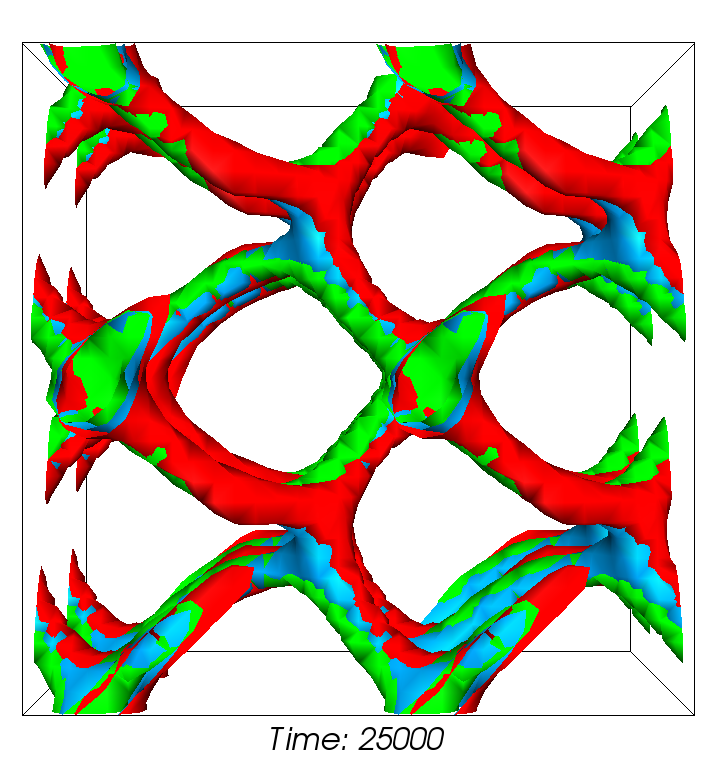}
\includegraphics[width=0.3\textwidth]{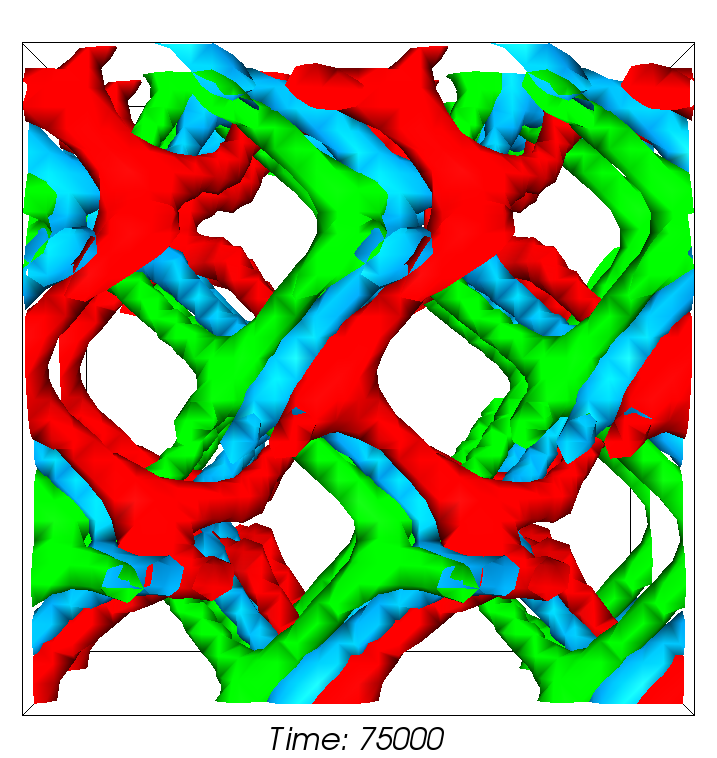}
\includegraphics[width=0.3\textwidth]{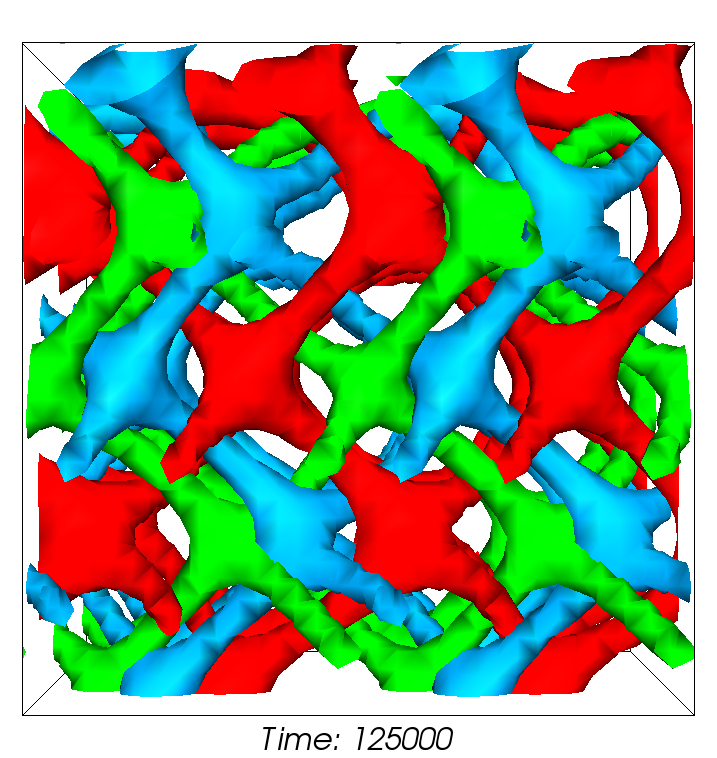}
\caption{BPII in shear flow for different thermodynamic parameters. The
snapshots show the evolution of the network with the flow direction into/out of
the page.  The reference state (blue, $\tau=-0.5, \kappa=2$), a high
temperature state (red, $\tau=0.5, \kappa=2$) and a low chirality state
(green, $\tau=-0.5, \kappa=1$) are shown; BPII is metastable in the later
two parameter sets.
The strain rate is $\dot{\gamma}=7.8125\ex{-5}$ LBU in all three cases.
The pictures show a central cubic section consisting of 8 unit cells cut
from a larger simulation with $n_y=4$.}
\label{fig2}
\end{figure*}

The speed of the network migration in the vorticity direction depends on the
thermodynamic state of the blue phase.
A comparison between the state from Fig.~\ref{fig1} with a
``high temperature'' state at $\tau=0.5, \kappa=2$ and a
``low chirality state'' at $\tau=-0.5, \kappa=1$ is shown in
Fig.~\ref{fig2}. Note that in both these states BPII is a metastable phase. 
Fig.~\ref{fig2} shows a sequence of initially overlapping
disclination networks which separate during the course of the simulation.
The network moves fastest in the high temperature state, is slower in the
reference state and slower still in the low chirality state.

\begin{figure*}[t]
\centering
\includegraphics[width=0.725\textwidth]{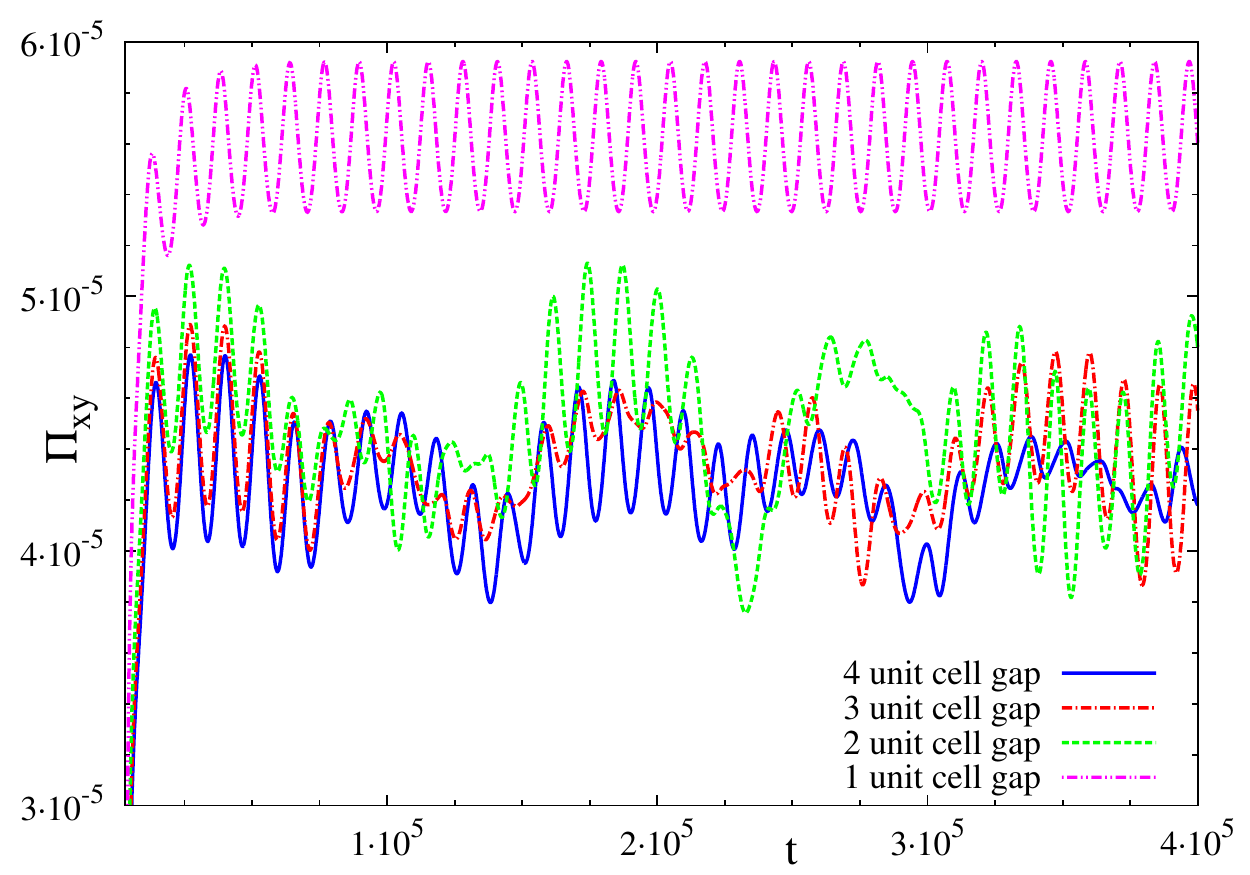}
\caption{BPII: ensemble-averaged shear stress $\langle \Pi_{xy}\rangle$
versus time in the reference state ($\tau=-0.5, \kappa=2$) for gaps ranging
from one to four unit cells.}
\label{fig3}
\end{figure*}

The total network displacement along the vorticity direction is more than
a unit cell in $10^5$ timesteps for the high temperature state.
Typical average and maximum values of the scalar order parameter are
$\bar{q}_s=0.338$, $q_{s,max}=0.383$ (low chirality state), 
$\bar{q}_s=0.206$, $q_{s,max}=0.320$ (high temperature state) and
$\bar{q}_s=0.304$, $q_{s,max}=0.408$ (reference state, in Fig.~\ref{fig1}).
The average scalar order parameter of the low chirality state is about
$10\%$ higher than that of the reference state. The  relative differerence
between the mean and the maximum value is about a factor $2.5$ smaller.
We propose that the speed of network migration is different because
permeation, which should be required for the stick-slip motion, may occur
more easily in states with less order (high temperature state) and gentler
spatial variation (low chirality state). Furthermore, the uneven stick-slip
nature of the network motion is possibly due to the anchoring at the
boundaries which prevents network displacements at the wall. 

Fig. \ref{fig3} shows the spatially averaged shear stress $\langle \Pi_{xy}\rangle$ versus time, in the reference state and for different gap sizes.
For gaps accommodating only one unit cell a regular periodic pattern emerges shortly after a transient start-up phase.
Every cycle is related to a break-up and reformation of the network.
The time-average in the steady state is about $\Pi_{xy}\simeq5.62\ex{-5}$
in LBU and yields an apparent viscosity
$\eta_{app}=(1 +\Pi_{xy}/\eta\dot{\gamma})\simeq 1.86$~LBU. This is
significantly higher than our baseline Newtonian viscosity $\eta = 0.8333$.
When the gap contains $n_y\ge2$ unit cells, the average stress shows
less regular oscillations, and fluctuates around a mean value which is
about $10-15\%$ smaller, roughly $\eta_{app}=1.65$ for $n_y=4$. The fact
that the average stress stabilises for $n_y\ge2$ indicates that bulk flow
behaviour is already reached for those thicknesses, but not in the single
unit cell case. Different shear rates lead to similar generic features.

\subsection{Blue Phase I}\label{sec_bpi}

\begin{figure*}[h]
\centering
\begin{minipage}[b]{0.3\textwidth}
\subfigure[]{\includegraphics[width=\textwidth]{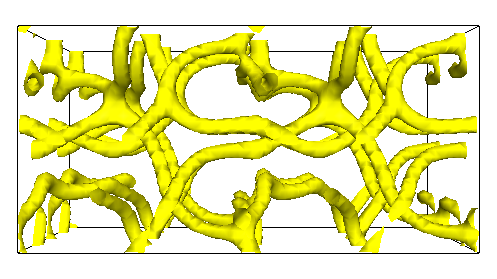}}
\subfigure[]{\includegraphics[width=\textwidth]{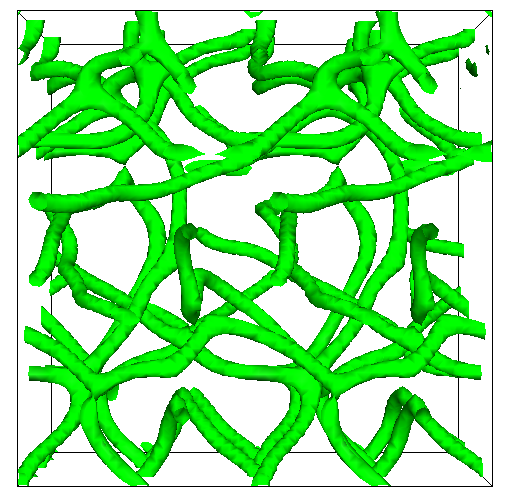}}
\end{minipage}
\subfigure[]{\includegraphics[width=0.3\textwidth]{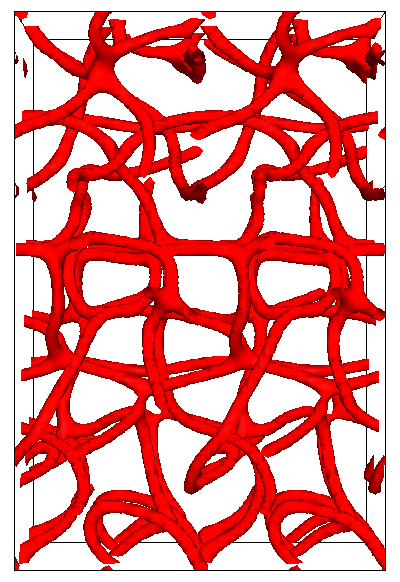}} 
\subfigure[]{\includegraphics[width=0.3\textwidth]{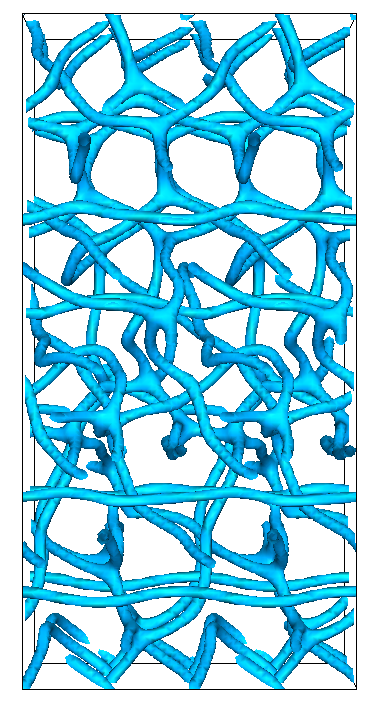}}
\caption{Confined BPI in shear flow at $\tau=-0.5, \kappa=1$ and gap sizes of one (a, yellow), two (b, green), three (c, red) and four (d, blue) unit cells. The pictures show the defect network along the flow direction at timestep $t=4\ex{5}$ at the same flow rate as above.}
\label{fig4}
\end{figure*}

We now compare and contrast the results obtained for BPII with those
pertaining to the rheology of BPI. In equilibrium, the topology
of the disclination network in BPI is very different from that of BPII.
The main difference is the presence in BPII of junction points where four
disclination lines of topological charge -1/2 meet; no such junctions
occur in BPI. We give representative results for BPI at $\tau=-0.5$
and $\kappa=1$, where it is the equilibrium phase.
Fig.~\ref{fig4} shows the disclination network at timestep $t=4\ex{5}$
for different gap sizes, with the viewing direction again chosen along
the flow direction.
For clarity we only show $n_x \times n_y \times n_z= 1 \times \{1,2,3,4\} \times 2$
unit cells.

\begin{figure*}[h]
\centering
\includegraphics[width=0.725\textwidth]{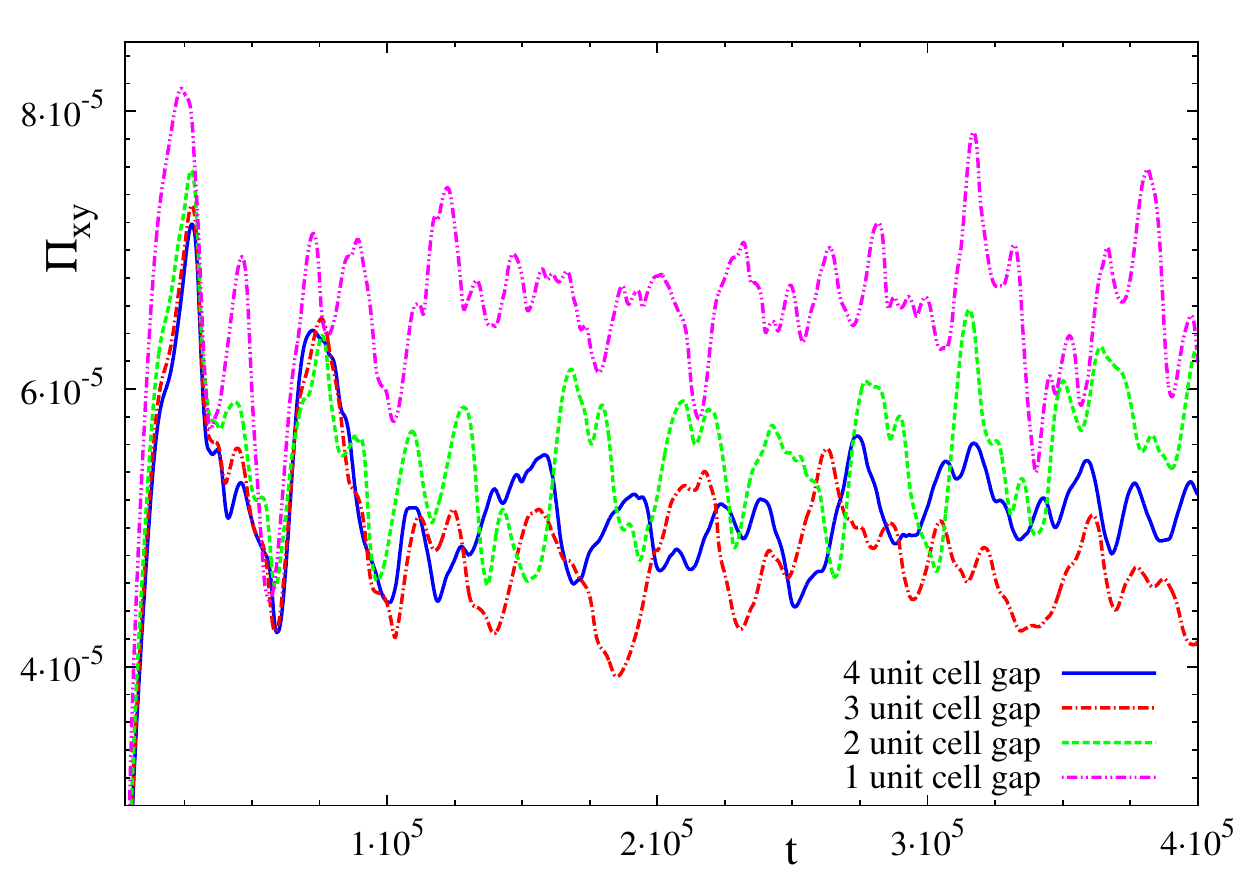}
\caption{BPI: ensemble-averaged shear stress $\langle \Pi_{xy}\rangle$
versus time in the reference state ($\tau=-0.5, \kappa=1$) for gaps
ranging from one to four unit cells.}
\label{fig5}
\end{figure*}

In contrast to BPII, the disclination network of BPI changes significantly
under a shear flow, and typical snapshots in Fig.~\ref{fig4} bear no
relation to the equilibrium BPI configuration. Furthermore, there is no sign of the ``stick-slip'' motion of the defects observed in BPII. While no regular cubic structure emerges under flow, it is interesting to note that the network develops a periodic modulation along the vorticity direction (horizontal in Fig.~\ref{fig4}). This periodicity arises for all gap sizes. Whether this feature is a consequence of frustration caused by the anchoring at the boundaries, or is a prelude to a slow ordering of the network under flow, we cannot yet tell.

The ensemble averages of the shear stress $\Pi_{xy}$ for different gap sizes
is shown in Fig.~\ref{fig5} and gives more evidence of the chaotic nature of
the flow of BPI. The magnitude of the average shear stress is comparable to
that seen in BPII; the fall in amplitude of the oscillations with increasing
gap size is also similar.

\section{Conclusions}

In conclusion, we have presented the first supra-unit cell simulations of
confined cubic blue phases in simple shear flow.
The rheological behaviour of BPI and BPII was found to be strikingly different.

Our simulations show that the disclination network of BPII continuously breaks and reforms while the network undergoes an affine deformation under shear. This process leads to an oscillatory shear stress in time, although the oscillations are only regular for gap sizes of a single unit cell across the flow gradient direction, where anchoring affects the physics of the system greatly, in agreement with recent simulations of quiescent BP networks \cite{Fukuda:2010a, Fukuda:2010b, Ravnik:2011b}.
For larger gap sizes (two and more unit cells) a ``stick-slip'' motion of
the network arises along the vorticity direction. The associated network
migration is not directly linked to mass flow, and is therefore somewhat
akin to permeation flow in cholesterics. The average speed of this network
migration depends on the thermodynamic parameters, and appears to be higher
for higher temperature and lower chirality. 

In contrast to BPII, BPI shows no regular break-up and relinking of the unit cells and its disclination network undergoes chaotic rearrangements under shear.
This behaviour was found for all gap sizes, and may be linked to the
strongly different topology in the equilibrium disclination patterns of
BPI and BPII. We will address the rheology of BPI in more detail in an
upcoming publication \cite{Henrich:2012}.
  
We hope that our results will stimulate further experimental work on 
the rheology of blue phases, similar to those reported e.g. in \cite{Negita:98}
but targeted to micron-sized samples, which would provide a direct test
of our simulations.

We finally note that the anchoring we have chosen does not lead to a conflict
with the bulk ordering of blue phases. In liquid crystal cells, it is 
customary to treat the surface so as to favour normal or planar anchoring
of the director field at the boundary. When used in very thin samples,
these boundary conditions lead to the formation of new disclination 
networks whose rheology it would be interesting to investigate~\cite{Fukuda:2010a,Fukuda:2010b,Tiribocchi:2011a}.

\ack
We acknowledge support by EPSRC Grants No. EP/E045316 and No. EP/E030173,
the MAPPER EU-FP7 project (grant no. RI-261507) and computing time on HECToR.
M.E.C. holds a Royal Society Research Professorship.


\section*{References}

\bibliography{lmcproc}

\begin{thebibliography}{10}

\bibitem{deGennes}
P.G. de~Gennes and J.~Prost.
\newblock {\em The Physics of Liquid Crystals (2nd Edition)}.
\newblock Clarendon Press, Oxford, 1993.

\bibitem{Grebel:1984}
S.~Shtrikman H.~Grebel, R.M.~Hornreich.
\newblock {\em Phys. Rev. A}, 30:3264, 1984.

\bibitem{Wright:1989}
D.C. Wright and N.D. Mermin.
\newblock {\em Rev. Mod. Phys.}, 61:385, 1989.

\bibitem{Henrich:2011a}
O.~Henrich, K.~Stratford, D.~Marenduzzo, and M.E. Cates.
\newblock {\em Phys. Rev. Lett.}, 106:107801, 2011.

\bibitem{Kikuchi:2002}
H.~Kikuchi, M.~Yokota, Y.~Hisakado, H.~Yang, and T.~Kajiyama.
\newblock {\em Nat. Mater.}, 1:64--68, 2002.

\bibitem{Coles:2005}
H.J. Coles and M.N. Pivnenko.
\newblock {\em Nature}, 436:997--1000, 2005.

\bibitem{Fukuda:2010a}
J.~Fukuda and S.~Zumer.
\newblock {\em Phys. Rev. Lett.}, 104:017801, 2010.

\bibitem{Fukuda:2010b}
J.~Fukuda and S.~Zumer.
\newblock {\em Liq. Cryst.}, 37:875--885, 2010.

\bibitem{Ravnik:2011b}
M.~Ravnik, G.P. Alexander, J.M. Yeomans, and S.~Zumer.
\newblock {\em Soft Matter}, 7:10144--10150, 2011.

\bibitem{Alexander:2008}
G.~P. Alexander and D.~Marenduzzo.
\newblock {\em EPL}, 81:66004, 2008.

\bibitem{Fukuda:2009}
J.~Fukuda, M.~Yoneya, and H.~Yokoyama.
\newblock {\em Phys. Rev. E}, 80:031706, 2009.

\bibitem{Henrich:2010a}
O.~Henrich, D.~Marenduzzo, K.~Stratford, and M.E. Cates.
\newblock {\em Phys. Rev. E}, 81:031706, 2010.

\bibitem{Castles:2010}
F.~Castles, S.M. Morris, E.M. Terentjev, and H.J. Coles.
\newblock {\em Phys. Rev. Lett.}, 104:157801, 2010.

\bibitem{Tiribocchi:2011}
A~Tiribocchi, G~Gonnella, D~Marenduzzo, and E~Orlandini.
\newblock {\em Soft Matter}, 7:3295, 2011.

\bibitem{Ravnik:2011a}
M.~Ravnik, G.P. Alexander, J.M. Yeomans, and S.~Zumer.
\newblock {\em Proc. Natl. Acad. Sci. USA}, 108:5188--5192, 2011.

\bibitem{Henrich:2010b}
O.~Henrich, K.~Stratford, D.~Marenduzzo, and M.E. Cates.
\newblock {\em Proc. Natl. Acad. Sci. USA}, 107:13212--13215, 2010.

\bibitem{Helfrich:1969}
W.~Helfrich.
\newblock {\em Phys. Rev. Lett.}, 23:372, 1969.

\bibitem{Marenduzzo:2006a}
D.~Marenduzzo, E.~Orlandini, and J.M. Yeomans.
\newblock {\em Phys. Rev. Lett.}, 92:188301, 2004.

\bibitem{Marenduzzo:2006b}
D.~Marenduzzo, E.~Orlandini, and J.M. Yeomans.
\newblock {\em J. Chem. Phys.}, 124:204906, 2006.

\bibitem{Rey:1996a}
A.D. Rey.
\newblock {\em Phys. Rev. E}, 53:4198--4201, 1996.

\bibitem{Rey:1996b}
A.D. Rey.
\newblock {\em J. Non-Newtonian Fluid Mech.}, 64:207--227, 1996.

\bibitem{Rey:2000}
A.D. Rey.
\newblock {\em J. Rheol.}, 44:855, 2000.

\bibitem{Rey:2002}
A.D. Rey.
\newblock {\em J. Rheol.}, 46:225, 2002.

\bibitem{Dupuis:2005}
A.~Dupuis, D.~Marenduzzo, E.~Orlandini, and J.M. Yeomans.
\newblock {\em Phys. Rev. Lett.}, 95:097801, 2005.

\bibitem{Zapotocky:1999}
M.~Zapotocky, L.~Ramos, T.C. Lubensky, and D.A. Weitz.
\newblock {\em Science}, 283:209, 1999.

\bibitem{Ramos:2002}
L.~Ramos, M.~Zapotocky, T.C. Lubensky, and D.A. Weitz.
\newblock {\em Phys. Rev. E}, 66:031711, 2002.

\bibitem{Beris:1994}
A.N. Beris and B.J. Edwards.
\newblock {\em Thermodynamics of Flowing Systems}.
\newblock Oxford University Press, 1994.

\bibitem{Alexander:2006}
G.P. Alexander and J.M. Yeomans.
\newblock {\em Phys. Rev. E}, 74:061706, 2006.

\bibitem{Denniston:2001}
C.~Denniston, E.~Orlandini, and J.M. Yeomans.
\newblock {\em Phys. Rev. E}, 63:056702, 2001.

\bibitem{Marenduzzo:2007}
D.~Marenduzzo, E.~Orlandini, M.E. Cates, and J.M. Yeomans.
\newblock {\em Phys. Rev. E}, 76:031921, 2007.

\bibitem{Denniston:2004}
C.~Denniston, D.~Marenduzzo, E.~Orlandini, and J.M. Yeomans.
\newblock {\em Philos. Trans. R. Soc. London, Ser. A}, 362:1745, 2004.

\bibitem{Henrich:2012}
O.~Henrich, K.~Stratford, D.~Marenduzzo, P.V. Coveney, and M.E. Cates.
\newblock Rheology of cubic blue phases.
\newblock {\em in preparation}.

\bibitem{Negita:98}
K~Negita.
\newblock {\em Liq. Cryst.}, 24:243--246, 1998.

\bibitem{Tiribocchi:2011a}
A~Tiribocchi, G~Gonnella, D~Marenduzzo, E~Orlandini, and F~Salvadore.
\newblock {\em Phys. Rev. Lett.}, 107:237803, 2011.

\end{thebibliography}
\bibliographystyle{unsrt}
\end{document}